%% file: icmc2025_paper_template.tex
\documentclass{article}
\usepackage{icmc2025_paper_template}
\usepackage{times}
\usepackage{ifpdf}
\usepackage{soul}
\usepackage[english]{babel}
\usepackage{multirow}
\usepackage[pdftex]{graphicx} 

\def\papertitle{A Real-Time Gesture-Based Control Framework}
\def\firstauthor{Mahya Khazaei
}
\def\secondauthor{Ali Bahrani
}
\def\thirdauthor{George Tzanetakis
}

\newif\ifpdf
\ifx\pdfoutput\relax
\else
   \ifcase\pdfoutput
      \pdffalse
   \else
      \pdftrue
  \fi
\fi

\ifpdf % compiling with pdflatex
  \usepackage[pdftex,
    pdftitle={\papertitle},
    pdfauthor={\firstauthor, \secondauthor},
    bookmarksnumbered, % use section numbers with bookmarks
    pdfstartview=XYZ % start with zoom=100% instead of full screen; 
   ]{hyperref}
  \graphicspath{{./figures/}}
  \DeclareGraphicsExtensions{.pdf,.jpeg,.png}

  \usepackage[figure,table]{hypcap}

\else % compiling with latex
  \usepackage[dvips,
    bookmarksnumbered, % use section numbers with bookmarks
    pdfstartview=XYZ % start with zoom=100% instead of full screen
  ]{hyperref}  % hyperrefs are active in the pdf file after conversion

  \usepackage[dvips]{epsfig,graphicx}
  \graphicspath{{./figures/}}
  \DeclareGraphicsExtensions{.eps}

  \usepackage[figure,table]{hypcap}
\fi

\hypersetup{
    colorlinks,%
    citecolor=black,%
    filecolor=black,%
    linkcolor=black,%
    urlcolor=black
}

\title{\papertitle}

 \threeauthors
   {0.5in}
   {\firstauthor}  {  Department  %of Computer Science 
   \\ University  of Victoria \\
     {\tt \href{mailto:maahhi@uvic.ca}{
     maahhi@uvic.ca
     }}}
   {\secondauthor} { %Independent Researcher\\
     {\tt \href{mailto:ali.bahrani519@gmail.com}{ali.bahrani519@gmail.com
     }}}
   {\thirdauthor}{Department% of Computer Science 
   \\ University of Victoria \\ %
     {\tt \href{mailto:gtzan@uvic.ca}{
     gtzan@uvic.ca
     }}}

\begin{document}
\capstartfalse
\maketitle
\capstarttrue
\begin{abstract}

We introduce a real-time, human-in-the-loop gesture control framework that can dynamically adapt audio and music based on human movement by analyzing live video input. By creating a responsive connection between visual and auditory stimuli, this system enables dancers and performers to not only respond to music but also influence it through their movements. Designed for live performances, interactive installations, and personal use, it offers an immersive experience where users can shape the music in real time.

The framework integrates computer vision and machine learning techniques to track and interpret motion, allowing users to manipulate audio elements such as tempo, pitch, effects, and playback sequence. With ongoing training, it achieves user-independent functionality, requiring as few as 50 to 80 samples to label simple gestures. This framework combines gesture training, cue mapping, and audio manipulation to create a dynamic, interactive experience. Gestures are interpreted as input signals, mapped to sound control commands, and used to naturally adjust music elements, showcasing the seamless interplay between human interaction and machine response.
%The abstract should be placed at the start of the top left column and should contain about 150-200 words. The abstract should be formatted in italic type (this has already been set in the abstract style).
\end{abstract}

\input{chapters/1-introduction} 
\input{chapters/2-relatedworks} 
\input{chapters/3-system}
\input{chapters/4-experiment}
\input{chapters/5-evaluation}
\input{chapters/6-conclusion}

%%%%%%%%%%%%%%%%%%%%%%%%%%%%%%%%%%%%%%%%%%%%%%%%%%%%%%%%%%%%%%%%%%%%%%%%%%%%%
%bibliography here
\bibliography{icmc2025_paper_template}

\end{document}

%% file: chapters/1-introduction.tex
\section{Introduction}\label{sec:introduction}
Sound control through motion and gesture detection is a research topic that bridges human-computer interaction, music, and sensory technologies. Interactive sound control systems aim to interpret physical movements and gestures, translating them into meaningful modifications of sound or music. This allows users to engage with audio naturally and intuitively through physical actions, overcoming the limitations of traditional interfaces, such as keyboards or knobs, that demand direct and deliberate input. Gesture-based sound control opens new possibilities for dynamic interaction and creative expression, both in performance and everyday applications \cite{visi2020interactivemachinelearningmusical}.

The primary objective of this project is to develop a real-time, gesture-based sound control framework that provides a seamless and intuitive interface for manipulating sound. Unlike conventional control systems that rely on tactile or manual input, our framework enables users to interact with sound environments using only their body movements. Our approach leverages advancements in computer vision and machine learning to detect and interpret gestures, making it possible to adjust sound features such as volume, pitch, and playback speed dynamically. Gesture-based sound control systems enable intuitive manipulation of sound through human movements, offering a dynamic alternative to traditional interfaces like keyboards or touchscreens. By leveraging technologies such as motion sensors, cameras, and physiological sensors, these systems capture continuous, fluid gestures, allowing real-time, nuanced sound control. This creates deeper interaction, where sound responds to body movements, enhancing expression, especially in live performances and interactive sound design \cite{2TheBodyAsMusicalInstrument-tanaka2019body}\cite{4ostersjo2016go}.

Despite challenges in mapping complex, multi-dimensional gesture data to sound parameters, advancements in technology and research into sonic affordances and human perception can inform the design of intuitive, expressive interfaces \cite{TowardsGesturalSonicAffordances}. These systems transcend musical performance, and can find applications in therapy, education, and multimedia:
\begin{itemize}
    \item \textbf{Therapeutic contexts:} Enhancing motor skills and emotional expression through active sound-movement interaction.
    \item \textbf{Educational settings:} Promoting immersive learning by integrating sound and movement for music and dance education.
    \item \textbf{Multimedia experiences:} Enriching interactive art, virtual reality, and gaming through synchronized gesture-sound control.
\end{itemize}

Rooted in embodied engagement, these systems merge sound, gesture, and motion, creating multi-sensory experiences that connect auditory and kinesthetic senses. By blending technology, art, and cognitive science, gesture-based systems open new avenues for creative expression and sensory interaction \cite{6godoy2010musical}. Code and supplemental material available at: \textit{GitHub link removed for anonymity} % \href{https://github.com/maahhi/realtime-gesturebased-audio-control/}{\textcolor {magenta}{GitHub}}

%% file: chapters/2-relatedworks.tex
\section{Related work}

\subsection{ Gesture and Body Movement Recognition Systems}

\subsubsection{Deep Learning for Gesture Recognition}

Recent advances in gesture recognition have benefitted from deep learning techniques. Key milestones include the use of Convolutional Neural Networks (CNNs) for pose estimation, such as Convolutional Pose Machines (CPMs), which iteratively refine predictions and address challenges such as occlusions and variations of poses \cite{34ConvolutionalPoseMachines}. Graph-based models such as spatial-temporal graph convolutional networks (ST-GCN) model human actions by treating skeleton sequences as graphs, capturing spatial and temporal dynamics without hand-crafted features \cite{35st-gcn}. Tools such as OpenPose enable 2D pose estimation in real-time of multiple people using Part Affinity Fields (PAFs), providing scalable and high-accuracy solutions \cite{36openPoseMultiPerson}. MediaPipe by Google further simplifies the prototyping pipelines for 2D and 3D pose detection, enabling high-accuracy real-time results \cite{37Mediapipe}.

\subsubsection{Ensemble-Driven Hand Pose Estimation}
Deep learning has also driven innovations such as the Five-Layer Ensemble CNN (5LENet) for 3D hand pose estimation. This approach breaks the task into sub-tasks, each estimating individual fingers, and merges the results for a complete 3D pose, addressing challenges such as self-occlusion and geometric ambiguities \cite{38five-layerEnsembleCNN}. Using the natural structure of the hand and dividing tasks into smaller parts, the system improves precision for applications such as VR, AR, and HCI \cite{39RNNHumanPoseEstimation}. Gesture recognition systems increasingly incorporate emotional context, as seen in research using motion capture data to interpret gestures for affective computing. Such approaches enhance HCI applications in VR, gaming, and robotics \cite{41kapur2005gesture}.

\subsubsection{Real-Time Gesture Recognition}

Real-time gesture recognition faces challenges such as latency, accuracy, and environmental variability. Applications in VR, gaming, and interactive systems require instant feedback. Techniques like Camshift and Haar-like classifiers enable real-time hand gesture tracking for virtual interactions but struggle with noise and lighting variability \cite{42Haar-like}. Sensor-based systems, such as those that use Kinect, leverage machine learning (e.g. Support Vector Machines) to classify gestures linked to emotional states, optimizing real-time performance on embedded devices \cite{43realtime-Kinect}.
 The ml5.js high-level JavaScript library is built on TensorFlow.js, and is designed to make machine learning accessible for creative coding \cite{ml5}. It provides pre-trained models for gesture recognition and body tracking, enabling real-time interactions in web applications with minimal setup. Its user-friendly API has been widely adopted in interactive media, education, and art-based projects. These advances collectively improve gesture recognition by improving precision, adaptability, and real-time interactivity in various applications.

\subsection{Real-time Sound Control Systems}

Real-time sound synthesis, especially in systems responding dynamically to environmental changes, faces challenges like low latency, resource optimization, and seamless interaction between virtual environments and sound. Research in this field tackles these issues using techniques such as GPU acceleration, efficient algorithms, and computational models to create adaptive soundscapes for gaming, live performances, and interactive applications.

One study utilized GPU-based recursive and non-recursive filters for real-time sound synthesis in virtual environments, overcoming the challenge of recursion on GPUs. This system efficiently handled dynamic audio requirements, synthesizing realistic collision sounds for immersive experiences \cite{realistic}. Another study recreated the sound of a historical acoustic wind machine using digital synthesis, optimizing performance through an Arduino-controlled engine and polyphony management for live performance scenarios \cite{mechanical-mapping}.

A third contribution introduced a double source-filter model to synthesize sounds of interactions like impacts and rolling. Using key parameters such as impact speed, the system dynamically adapted to changes in the environment while maintaining low computational demands, suitable for gaming applications \cite{Morphing-Solids-Interactions}. Research on footstep sound synthesis for VR utilized physical models to generate realistic audio feedback based on ground reaction forces, enhancing immersion without requiring sensor-embedded footwear \cite{footsteps-for-VR}. Lastly, a real-time vocal harmonization system for live performances employed frame-based processing with the YIN pitch detection algorithm and PSOLA for pitch shifting, ensuring low latency and high-quality sound \cite{backvocal-harmony}. These works highlight advancements in hardware optimization, algorithm design, and system integration, significantly improving real-time sound synthesis and enriching user experiences in gaming, VR, and live performances.

\subsection{Gesture-based Interactive Sound Mapping systems}

The Wekinator \cite{wekinator}, created by Rebecca Fiebrink, is a machine learning software designed for real-time, interactive applications, enabling users to create musical instruments, sound controllers, and other creative tools. Tailored for artists and musicians, it simplifies training machine learning models, allowing non-experts to explore gesture, sound, and image recognition. Wekinator supports various input devices like cameras, microphones, and sensors, translating them into data for algorithms that produce outputs such as sounds or visuals. It integrates with software like Max, Pure Data, and Processing and offers multiple algorithms, including k-nearest neighbors and neural networks. With its focus on real-time interactivity and ease of use, Wekinator is a popular tool for creative technology projects, removing coding barriers for artists and designers and an important influence in the design and development of the proposed system. 

In the Continuous Interaction with a Smart Speaker (CISS) system  the authors introduce a novel method for controlling smart speakers through mid-air hand gestures via low-dimensional embeddings of dynamic hand pose \cite{continuousinteractionsmartspeaker}. The system utilizes the Google's MediaPipe Hands to extract 21 hand landmarks from video frames, which are then embedded into a two-dimensional pose space using an autoencoder. This low-dimensional embedding facilitates intuitive interaction by mapping hand poses to corresponding music track profiles. A PointNet-based model is employed to classify gestures, enabling users to control music playback and explore music spaces seamlessly. By jointly optimizing the autoencoder with the classifier, the system enhances gesture discrimination, allowing users to select different musical moods through variations in hand pose. This combination ensures efficient and responsive control of smart speaker functions through
 dynamic hand gestures. Our proposed system also utilizes the Google  MediaPipe framework with a focus on interfacing with sound and music creation software and the Python ML ecosystem. 
 
 In the paper Mugeetion\cite{mugeetion}, the authors present a novel musical interface that captures users’ facial gestures to influence musical features in real-time. The system utilizes FaceOSC software to track facial gestures and employs the Facial Action Coding System (FACS) to interpret these gestures as emotional states. By mapping specific facial action units (AUs) to corresponding musical parameters, Mugeetion enables users to control sound generation through their facial expressions, creating an emotion-driven musical experience. This approach leverages computer vision techniques for real-time facial expression detection, allowing for dynamic modulation of sound based on the performer’s emotional state.
 
 In their 2020 work, " Towards assisted interactive machine learning(AIML)", Federico Ghelli Visi and Atau Tanaka introduce a system that employs reinforcement learning to develop adaptive gesture-sound mappings \cite{visi2020towards}. This approach allows musicians to interactively train the system, enabling it to learn and refine the associations between specific gestures and corresponding sounds over time. By integrating reinforcement learning, the system adapts to the performer’s unique gestures, facilitating a personalized and responsive musical experience. This method contrasts with traditional static mappings by offering dynamic, real-time adaptability, thereby enhancing expressive potential in musical performances. Table \ref{table:comparison_accuracy_gesture_python} summarizes these different systems for gesture control.

\begin{table}[h]
    \centering
    \caption{Comparison of Systems Based on Accuracy, Gesture Types Supported, and Python Compatibility}
    \label{table:comparison_accuracy_gesture_python}
    \resizebox{\columnwidth}{!}{%
        \begin{tabular}{|p{2.5cm}|p{2.5cm}|p{2.5cm}|p{2.5cm}|}
            \hline
            \textbf{System} & \textbf{Accuracy} & \textbf{Gesture Types Supported} & \textbf{Python Compatibility} \\
            \hline
            Proposed & Hypothetically 85-90\% based on setup & Full-body and hand gestures & Yes \\
            \hline
            CISS & Not specified & Mid-air hand gestures & Yes \\
            \hline
            Mugeetion & Not specified & Facial movements, emotion-driven gestures & No \\
            \hline
            Wekinator & Dependent on user configuration and training iterations & Multiple gesture types based on user-defined inputs & No \\
            \hline
            AIML & Dependent on RL training and adaptive learning performance & Hand and body gestures with reinforcement-based adjustments & Yes \\
            \hline
        \end{tabular}%
    }
\end{table}

%% file: chapters/3-system.tex
\section{system Description}
The proposed system processes live video streams to alter sound based on body
 movements and postures, providing an interactive, real-time experience. Body detec
tion and landmark extraction are handled by Google’s MediaPipe framework, which
 identifies key body, hand, and facial landmarks necessary for interpreting gestures.
 MediaPipe runs within the Max/MSP programming environment through an emulation setup, 
 leveraging Max MSP’s real-time processing strengths in multimedia
 applications. Since the most effective machine learning tools and models are avail
able in the Python environment, we used the OSC (Open Sound Control) protocol \cite{OSC} to
 bridge Max/MSP and Python. OSC ensures fast, structured communication between
 these environments, allowing gesture data extracted in Max MSP to be processed by
 machine learning models in Python for adaptive sound synthesis. This setup enables
 a dynamic response between visual input and audio output, enhancing the system’s
 real-time interactivity and interfacing the large state of the art ML ecosystem of the
 Python language.

 \subsection{System architecture}

The system architecture integrates Max MSP and Python to enable real-time gesture-based sound control, with Open Sound Control (OSC) facilitating communication between the two environments. Max MSP captures video from a live camera feed, which is processed by MediaPipe to extract body landmarks representing key positions of the user. These landmarks are formatted as JSON and sent to Python via OSC.

The Python module operates differently depending on the phase: training or performance. In the training phase, the incoming landmark data is saved and used to build and train machine learning models for gesture classification. During the performance phase, the trained (in Python) models are used to classify incoming landmarks in real-time. Once a gesture is identified, Python sends a signal back to Max/MSP, indicating the recognized gesture class.

\begin{figure}[h]
    \centering
    \includegraphics[width=1\linewidth]{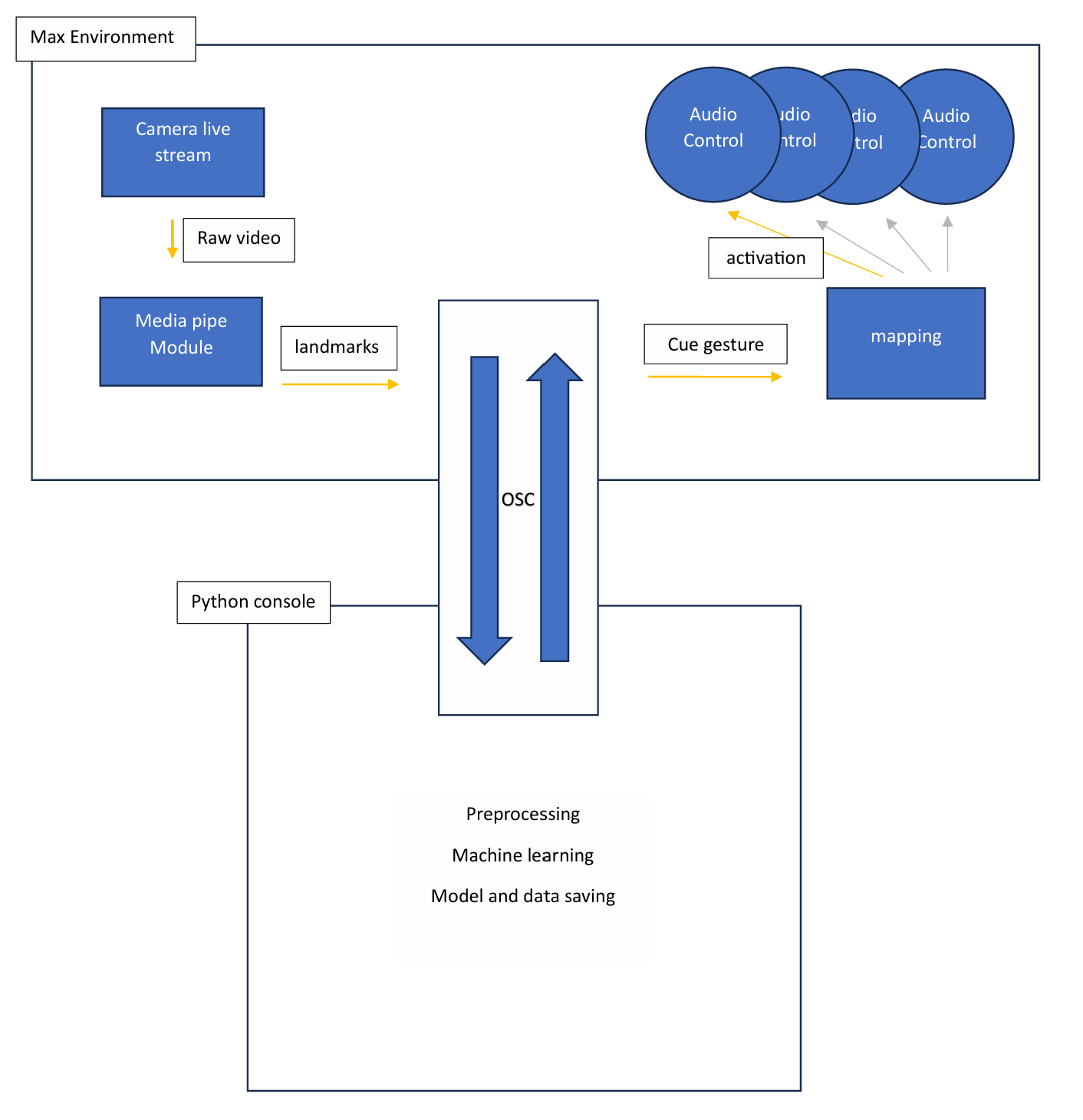} 
    \caption{System Data Flow Architecture}
    \label{fig:arch-diagram}
\end{figure}

The Max MSP part of the system allows users to map gestures to specific sound control modules, defining how each gesture influences audio manipulation. In the performance environment, these mappings are applied to enable live, dynamic sound control based on user-defined associations. The system's modular design ensures efficient data flow, with Max/MSP handling video capture, landmark extraction, and audio manipulation, while Python manages machine learning processes, classification,  and storage management. This workflow provides flexibility, adaptability, and scalability, making the architecture suitable for interactive sound applications in both experimental and performance contexts. It is shown in figure \ref{fig:arch-diagram}. 

\subsection{System Components}
\subsubsection{Audio Manipulation}
Max/MSP enables users to build interactive, real-time systems by visually connecting objects that represent various functions, from sound synthesis to video manipulation. Its modular and extensible design makes it adaptable to a wide range of creative and technical projects.

Max/MSP is optimized for low-latency, real-time multimedia processing, making it ideal for applications like interactive music generation or responsive sound control. A key feature is its ability to apply audio effects to audio files during playback, a complex task that Max/MSP simplifies through its intuitive visual interface. This functionality is particularly valuable for projects requiring dynamic audio manipulation in real-time. Compared to alternatives like PyGame, which is more suited for simpler 2D game graphics, Max/MSP delivers faster response times and superior performance on standard hardware, making it the preferred choice for intricate sound and multimedia applications.

\subsubsection{Body Landmark Extraction}

MediaPipe, developed by Google, is a versatile framework for building real-time machine learning pipelines for video, audio, and sensor data. Widely used for applications like object detection, hand and body tracking, and face detection, it offers pre-built solutions for specific use cases. For instance, MediaPipe Hands tracks 21 hand landmarks for gesture recognition, Face Mesh detects 468 facial landmarks for detailed expression analysis, and Pose provides 33 body landmarks for activity tracking and includes built-in hand gesture detection. To integrate MediaPipe into Max MSP, developers use jweb-mediapipe—a JavaScript-based adaptation originating from Lintang Wisesa's MediaPipe in JavaScript project. This library allows Max MSP to leverage MediaPipe’s body and facial landmark detection for real-time, low-latency multimedia interactions, ideal for gesture-driven applications. Our system utilizes MediaPipe for gesture recognition, processing 3D coordinates (x, y, z) of detected landmarks. For hand gestures, all landmarks are used to ensure precise tracking. For body pose recognition, only relevant landmarks are selected, with the nose position serving as a stable anchor for head orientation while excluding points like face landmarks which are unnecessary for movement analysis.

\begin{figure}[h!]
    \centering
    \includegraphics[width=1\linewidth]{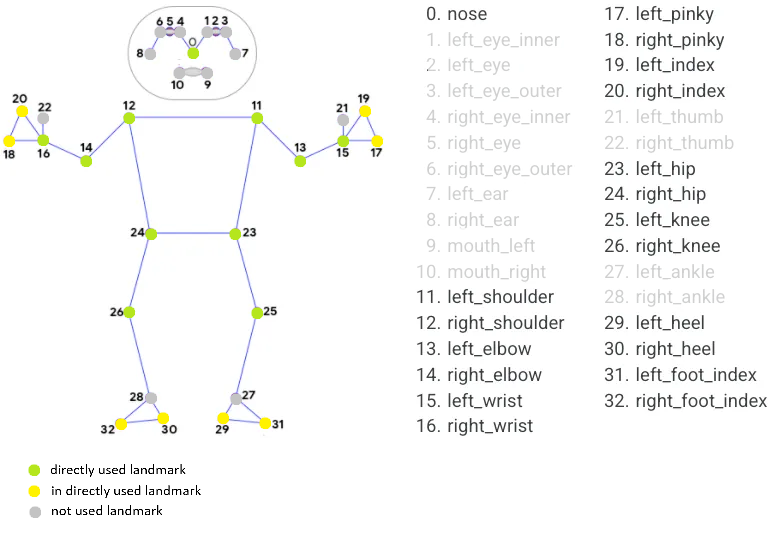}
    \caption{Mediapipe body pose landmarks and feature selection}
    \label{fig:body-landmarks}
\end{figure}

\subsubsection{Bridging Real-Time Communication}
Open Sound Control (OSC) is a lightweight communication protocol designed for real-time, low-latency data exchange between multimedia applications. Initially developed in the 1990s for music applications, OSC is now widely used in interactive installations, gaming, and virtual reality due to its flexibility and efficiency. It transmits structured messages containing address patterns and arguments, enabling dynamic communication across devices and platforms. This makes it ideal for synchronizing sound, video, and other media in complex, interactive systems.

OSC is particularly suited for sending data from Max/MSP to Python in real-time projects. For example, gesture data from MediaPipe in Max/MSP can be efficiently transmitted to Python for machine learning processing using frameworks like PyTorch and TensorFlow. OSC’s compatibility with these frameworks allows seamless integration, enabling the use of pre-trained models and custom implementations for real-time gesture data processing. This modular and extendable approach facilitates experimentation and adaptability. 

\subsubsection{Machine Learning Mode}
In our initial experiments, we evaluated several classifiers, including Random Forest, Gradient Boosting, and Multilayer Perceptron (MLP), for classifying body gestures based on body landmarks. While Random Forest and Gradient Boosting performed reasonably well, the MLP outperformed them in accuracy and consistency, proving to be the most effective choice. The MLP excels in body gesture classification due to its ability to capture complex, non-linear relationships in high-dimensional data. Its fully connected layers effectively identify dependencies among spatial configurations of body landmarks, enabling precise differentiation of subtle gesture variations. Furthermore, MLPs are adaptable, requiring minimal computational overhead and preprocessing. Unlike convolutional networks (CNN), they can directly handle flattened data structures, such as body landmark coordinates, efficiently without having to convert them to a matrix representation. Their computational efficiency and strong performance make MLPs ideal for real-time gesture recognition systems, balancing speed and accuracy effectively.

%% file: chapters/4-experiment.tex
\section{Experiments}

Our goal is to explore how performers can use distinct gestures to manipulate sound in real time, enabling new levels of creative expression. To investigate this, we designed two test scenarios simulating performance conditions to identify the technical requirements for accurate gesture recognition.

The process includes distinct phases: preparation, training, and evaluation. In the training phase, performers execute their primary gesture in a metronome-guided sequence, helping the system distinguish it from incidental movements. During evaluation, the system's gesture recognition accuracy is assessed, and users can refine the model by adding additional instances or corner cases to improve robustness for live use.

Once trained and validated, performers map gestures to specific sound actions in Max MSP for real-time control. The setup enables adjustments to playback timing, pitch, gain, and effects based on detected gestures. This methodical approach validates gesture-based sound control as a novel tool for interaction in performance contexts.

\subsection{Experimental Overview}
The primary goal of this experiment is to validate the feasibility and effectiveness of a real-time gesture-controlled sound synthesis system. It aims to demonstrate that body-tracked gestures (e.g., via MediaPipe) can responsively modulate audio output in real time. Key objectives include evaluating system responsiveness (low latency), gesture detection accuracy, and reliability in maintaining continuous sound control. Additionally, it assesses the integration of components (Max MSP, MediaPipe, and Python via OSC) and evaluates usability, focusing on intuitive control and adaptability to diverse gestures. These efforts aim to establish the system's practicality for dynamic environments like live performances and interactive installations.

Two scenarios were developed to address real-world performance challenges.
\begin{enumerate}
    \item {\bf Dance Performance Scenario}: This scenario addresses synchronization issues in performances that utilize pre-recorded tracks. Here, performers can trigger the music system to continue from specific points during the track, enabling dialogue and movements to align with musical moments, creating a cohesive and effective performance.
    \item {\bf Real-Time Sound Control Scenario}: This setup enables performers to control sound elements—adjusting volume, pitch, and gain of drums, instruments, or vocals—using designated gestures. Unlike tools like Wekinator, which focus on creating digital instruments, this scenario enhances real-time modifications to existing tracks.
\end{enumerate}

\subsection{Training phase}

The training phase for the machine learning model is designed to help users intuitively provide effective samples in a fun and engaging way. The system ensures users have ample time to prepare before making gestures and eliminates the need to press any buttons for each entry. Here's how it works:
\begin{enumerate}

\item \textbf{Setting the Tempo:} Users start by selecting either a plain beat or the main audio track as their tempo.
\item \textbf{Sample Count Selection:} Users choose the number of samples to provide, with 4 or 8 samples recommended initially.
\item \textbf{Countdown \& Cue Readiness:} A visual countdown ("3, 2, 1") and plain beats signal users to prepare for the first cue gesture.
\item \textbf{Sample Collection:} A green light and distinct beat signal the start of sample recording. A sequence of three yellow beats followed by one green beat repeats, with each green beat prompting the user to perform the "cue gesture." Yellow beats allow neutral or other movements to help distinguish the cue gesture during real-time performance.
\item \textbf{Confirmation:} Once samples are collected, users review and confirm their accuracy, discarding any incorrect samples to ensure precise data for training, improving the model’s performance.

\end{enumerate}

\subsubsection{Machine learning training}
During each green light moment, a snapshot of body position landmarks detected by the MediaPipe module in the Max MSP environment is captured via a live stream from the user’s laptop camera. These landmarks represent the user’s posture at that moment and are sent as JSON objects to the Python environment via the OSC protocol. Snapshots during green light moments are labeled as the “main class” for the primary gesture, while those during yellow beats are labeled as the “other” class for non-primary movements. This labeled dataset is used to train machine learning models to classify gestures accurately. Data preprocessing standardizes and scales inputs, ensuring consistency. Each landmark includes (x, y, z) coordinates, but only x and y are used to simplify the process while retaining spatial information.

\subsection{Training Evaluation}

After training, the model processes the data to learn the intended gesture and provides evaluation metrics to assess its recognition accuracy. If accuracy is insufficient, additional samples are collected, and training is repeated to improve performance.

The user can test the model in a live environment to observe its responsiveness and ensure accurate interpretation of gestures. Similar to platforms like Wekinator, this system allows adding corner cases to address model weaknesses, enhancing precision across various conditions.

Once trained, gestures can be assigned custom names for easy identification. The system generates a model file and a scaler file for saving the trained model and preprocessing parameters, enabling reuse without retraining. Users can test saved gestures to ensure the system reliably distinguishes between different cue gestures.

\subsection{Mapping}
 The gestures saved during the final stage of the training phase are now fully recognizable by the system. At this point, the user can select from various scenarios, each tailored to different performance contexts. Based on the chosen scenario, the user is prompted to provide specific information that links each gesture to the desired audio control functions.

 \subsection{Action Phase}
 The system is now ready to recognize the cue gesture and trigger the corresponding action. It uses an observation window—a defined time frame to monitor body landmarks and estimate the probability of detecting the gesture. Once the gesture is confidently identified, the assigned action is executed. The observation window can remain open for continuous responsiveness or be shortened in scenarios like dance performances to minimize misinterpretation and improve accuracy in detecting intended gestures.

\subsubsection{ Dance Performance Scenario}
In the first scenario, the user sets an observation window with start and end times to focus on detecting the cue gesture, improving accuracy. A specific cue moment in the soundtrack, like a bang sound at time t, is aligned with the performer’s gesture.

The user also configures a latency tolerance window to handle timing variations. During playback, the system monitors for the cue gesture within the observation window. If detected early, the system triggers the bang sound at the designated moment (t). If the gesture isn’t detected, the observation window extends by the latency tolerance period, allowing for slight delays. If no gesture is detected by the end, playback proceeds from the bang sound. This process is designed to integrate seamlessly with traditional dance performances, maintaining timing and choreography.

\subsubsection{ Real-time sound control Scenario}
This scenario provides flexible control over sound elements by linking gestures to both directional commands (e.g., increase or decrease) and unit values. For example, pointing upward may increase volume by one unit, while pointing downward decreases it. More dynamic gestures, like raising a whole hand, could adjust volume by five units for a pronounced effect, enabling subtle or dramatic sound changes based on gesture expressiveness.

The system scans for gestures at each beat, capturing brief landmark snapshots and responding within milliseconds to recognize saved gestures. This ensures quick, reliable recognition while minimizing accidental commands.

Before activation, the audio track is divided into stems (e.g., drums, vocals). Specific gestures can adjust the gain of each stem independently, allowing for custom arrangements and nuanced soundscapes. This approach combines directional commands, adjustable units, and independent stem control, giving performers real-time creative freedom for dynamic audio manipulation.

%% file: chapters/5-evaluation.tex
\section{Evaluation}
The system was successfully implemented and tested with two users performing multiple gestures. Feedback time was measured to ensure real-time responsiveness. A comparative analysis with MediaPipe’s hand gesture recognizer assessed accuracy and reliability, validating the system's effectiveness for dynamic applications.

\subsection{Participants analytics}

In this study, two users participated in a training phase involving four gestures: raising the right/left hand and right/left leg. Table \ref{tab:sampless} details the sample count for each gesture class. User A trained for 6 minutes, and User B for 5 minutes, with an average sample time of 2.6 seconds, including interaction and rest. Training continued until the system provided accurate, real-time feedback for each gesture. Model precision, recall, F1 score, and accuracy are presented in Tables \ref{table:A-classification_report} and \ref{table:B-classification_report}.

\begin{table}[h!]
\centering
\begin{tabular}{|c|c|c|}
\hline
\textbf{Gesture} & \textbf{Participant A} & \textbf{Participant B} \\ \hline
Right Hand Up            & 58            & 50            \\ \hline
Right Leg Up            & 36            & 50           \\ \hline
Left Hand Up            & 36            & 60            \\ \hline
Left Leg Up            & 24            & 60            \\ \hline
\end{tabular}
\caption{Number of Samples for Each Gesture by Participants A and B}
\label{tab:sampless}
\end{table}

\begin{figure}[h!]
    \centering
    \includegraphics[width=1\linewidth]{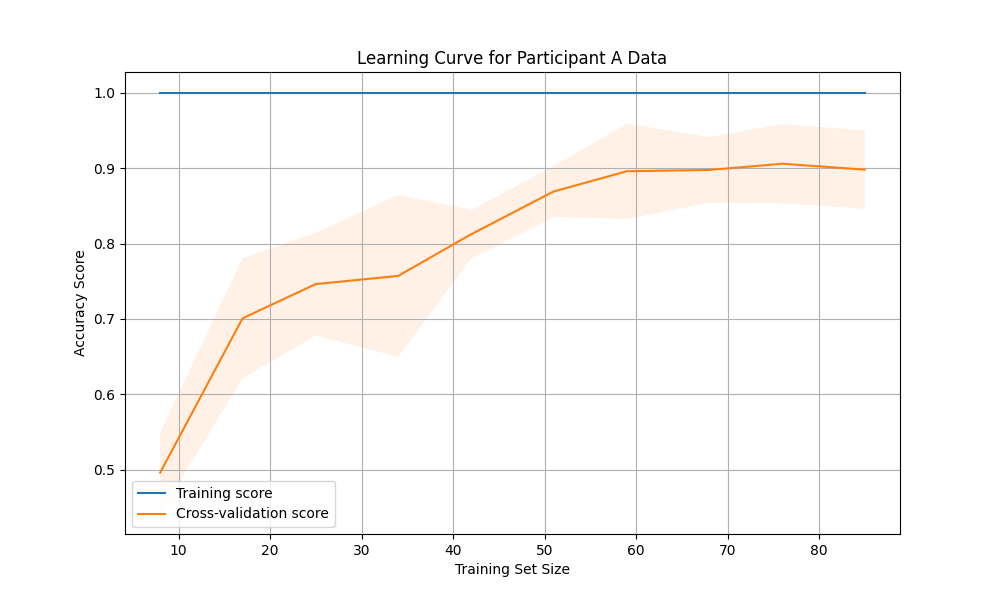}
    \caption{user A learning curve}
    \label{fig:A-learningcruve}
    
    \vspace{1em} % Add vertical space between the images

    \includegraphics[width=1\linewidth]{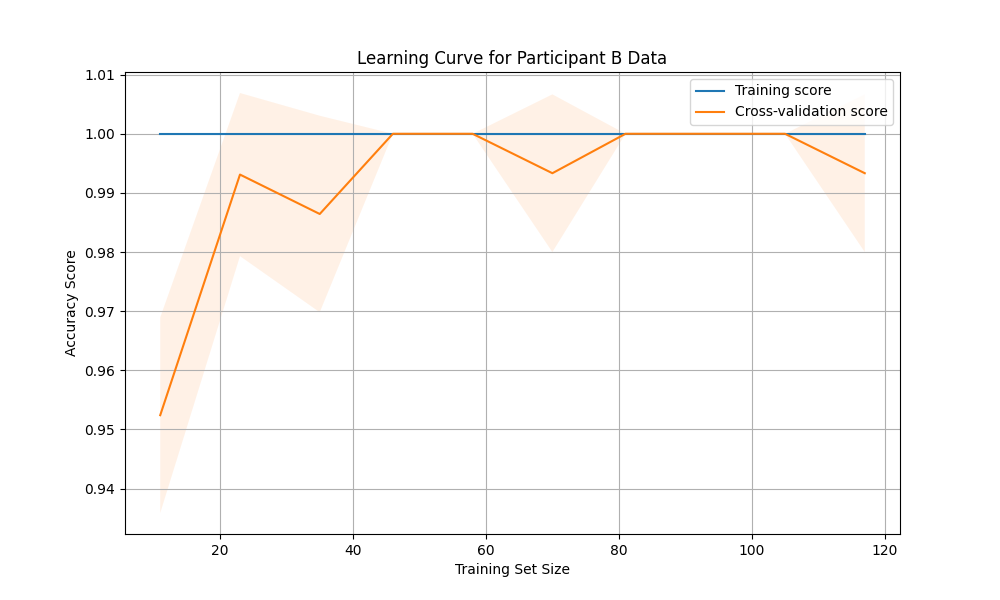}
    \caption{user B learning curve}
    \label{fig:B-learningcruve}
\end{figure}

Accuracy stabilized after ~60 samples, as shown in Figures \ref{fig:A-learningcruve} and \ref{fig:B-learningcruve}, though ~50 samples were sufficient for reliable classification. Cross-user testing revealed lower performance when User A’s model was tested with User B’s data, highlighting the importance of user-specific training to improve convergence and real-time accuracy. A generalized model trained on diverse datasets can achieve broader applicability if the features (e.g., landmarks) are not body-specific. Comparative performance data can be found in Table \ref{table:cross-user-performance}.

\begin{table}[h!]
\centering
\begin{tabular}{|c|c|c|c|c|}
\hline
\textbf{Class}       & \textbf{Precision} & \textbf{Recall} & \textbf{F1-Score} & \textbf{Support} \\ \hline
Right Hand Up        & 1.00               & 1.00            & 1.00              & 19               \\ \hline
Left Hand Up         & 1.00               & 0.89            & 0.94              & 9                \\ \hline
Right Leg Up         & 0.80               & 1.00            & 0.89              & 12               \\ \hline
Left Leg Up          & 1.00               & 0.71            & 0.83              & 7                \\ \hline
\textbf{Accuracy}    & \multicolumn{3}{c|}{0.94}            & 47               \\ \hline
\textbf{Macro Avg}   & 0.95               & 0.90            & 0.92              & 47               \\ \hline
\textbf{Weighted Avg}& 0.95               & 0.94            & 0.94              & 47               \\ \hline
\end{tabular}
\caption{Classification Report for participant A based on validation dataset}
\label{table:A-classification_report}
\end{table}

\begin{table}[h!]
\centering
\begin{tabular}{|c|c|c|c|c|}
\hline
\textbf{Class}       & \textbf{Precision} & \textbf{Recall} & \textbf{F1-Score} & \textbf{Support} \\ \hline
Right Hand Up                  & 1.00               & 1.00            & 1.00              & 16               \\ \hline
Left Hand Up                  & 1.00               & 0.84            & 0.91              & 19               \\ \hline
Right Leg Up                  & 0.82               & 1.00            & 0.90              & 14               \\ \hline
Left Leg Up                  & 1.00               & 1.00            & 1.00              & 14               \\ \hline
\textbf{Accuracy}    & \multicolumn{3}{c|}{0.95}            & 63               \\ \hline
\textbf{Macro Avg}   & 0.96               & 0.96            & 0.95              & 63               \\ \hline
\textbf{Weighted Avg}& 0.96               & 0.95            & 0.95              & 63               \\ \hline
\end{tabular}
\caption{Classification Report for participant B based on validation dataset}
\label{table:B-classification_report}
\end{table}

\subsection{Real-time performance}
To evaluate real-time responsiveness, we measured the total cycle time from capturing body landmarks to triggering audio output. This process, including movement capture, landmark transmission, cue recognition, and audio response, consistently occurs within 0.2 seconds. This sub-200ms latency meets the application's real-time requirements, providing immediate feedback for seamless and intuitive audio interaction.

\subsection{ Comparative Classification Evaluation}

To evaluate our gesture classification system against an established benchmark, we compared it to MediaPipe’s hand gesture recognition model. The experiment focused solely on hand landmarks and categorized gestures into three classes: Open Palm, Closed Fist, and other hand gestures.

Using MediaPipe’s gesture labeling as the ground truth, we assessed our model's accuracy. The results showed a 90\% performance similarity to MediaPipe, with high precision and recall across most categories, demonstrating the effectiveness of our training system for hand gesture classification.

\begin{table}[h!]
\centering
\begin{tabular}{|c|c|c|}
\hline
\textbf{Class}       & \textbf{F1-Score} & \textbf{Support} \\ \hline
Open Palm            & 0.96              & 64               \\ \hline
Close Fist           & 0.89              & 71               \\ \hline
Others               & 0.89              & 107              \\ \hline
\textbf{Accuracy}    & \multicolumn{2}{c|}{0.91}            \\ \hline
\textbf{Macro Avg}   & 0.91              & 242              \\ \hline
\textbf{Weighted Avg}& 0.91              & 242              \\ \hline
\end{tabular}
\caption{Comparison of Our Model with MediaPipe Hand Gesture Recognition System}
\label{table:mediapipe_comparison}
\end{table}

%% file: chapters/6-conclusion.tex
\section{CONCLUSION }
A real-time, gesture-based sound control system is introduced, offering performers an intuitive, customizable interface for manipulating sound through body movements. The system bridges human movement with machine learning to create an adaptable and expressive tool, accessible to users with or without technical expertise.

Key contributions include enabling performers to customize system parameters for greater accuracy and responsiveness, empowering personalized interaction and creative expression. The system’s ability to generalize across diverse users through a comprehensive dataset highlights its potential for broad applications, from live performances to therapeutic environments. By integrating machine learning frameworks like PyTorch and TensorFlow, it simplifies advanced sound manipulation, making it accessible even to non-technical users.

However, limitations such as sensitivity to environmental factors, initial training time, and challenges in handling complex gestures remain. Addressing these issues through enhanced environmental robustness, adaptive learning, and advanced motion detection could improve the system’s performance and expand its applicability. Future work may explore applications in therapy and education, leveraging Python's ecosystem for real-time learning and integrating cutting-edge gesture recognition technologies for more intricate movement tracking.

This research demonstrates the potential of combining gesture recognition and machine learning for sound control, offering performers a dynamic, intuitive tool for creative expression. By uniting technology and art, it paves the way for more immersive and inclusive multimedia experiences.

\begin{table}[h!]
\centering
\small % Reduce font size for the table
\begin{tabular}{|c|c|c|c|c|c|}
\hline
\textbf{Test} & \textbf{Class}       & \textbf{Precision} & \textbf{Recall} & \textbf{F1-Score} & \textbf{Support} \\ \hline
\multirow{7}{*}{A on B} 
                       & Right Hand Up        & 0.71               & 0.88            & 0.78              & 58               \\ \cline{2-6}
                       & Left Hand Up         & 0.64               & 0.81            & 0.72              & 36               \\ \cline{2-6}
                       & Right Leg Up         & 0.88               & 0.58            & 0.70              & 36               \\ \cline{2-6}
                       & Left Leg Up          & 0.31               & 0.17            & 0.22              & 24               \\ \cline{2-6}
                       & \textbf{Accuracy}    & \multicolumn{3}{c|}{0.68}            & 154              \\ \cline{2-6}
                       & \textbf{Macro Avg}   & 0.63               & 0.61            & 0.60              & 154              \\ \cline{2-6}
                       & \textbf{Weighted Avg}& 0.67               & 0.68            & 0.66              & 154              \\ \hline
\multirow{7}{*}{B on A} 
                       & Right Hand Up        & 1.00               & 0.98            & 0.99              & 50               \\ \cline{2-6}
                       & Left Hand Up         & 0.98               & 1.00            & 0.99              & 50               \\ \cline{2-6}
                       & Right Leg Up         & 1.00               & 0.97            & 0.98              & 60               \\ \cline{2-6}
                       & Left Leg Up          & 0.96               & 1.00            & 0.98              & 50               \\ \cline{2-6}
                       & \textbf{Accuracy}    & \multicolumn{3}{c|}{0.99}            & 210              \\ \cline{2-6}
                       & \textbf{Macro Avg}   & 0.99               & 0.99            & 0.99              & 210              \\ \cline{2-6}
                       & \textbf{Weighted Avg}& 0.99               & 0.99            & 0.99              & 210              \\ \hline
\end{tabular}
\caption{Cross-User Performance of Models: User A Data Tested with Model Trained on User B Data (A on B) and Vice Versa (B on A)}
\label{table:cross-user-performance}
\end{table}